\documentclass[a4paper,twocolumn,english,prl,superscriptaddress,showpacs,longbibliography,fixfloat]{revtex4-1}
\usepackage[latin9]{inputenc}
\usepackage{color}
\usepackage{amsmath}
\usepackage{amssymb}
\usepackage{graphicx}

\makeatletter

\pdfpageheight\paperheight
\pdfpagewidth\paperwidth

\usepackage{xcolor}
\definecolor{darkblue}{rgb}{0.1,0.2,0.6} 
\definecolor{lightblue}{rgb}{0.1,0.1,1.0}
\definecolor{darkred}{rgb}{0.8,0.1,0.2}
\usepackage[colorlinks,citecolor=lightblue,linkcolor=darkblue,urlcolor=lightblue] {hyperref}
\renewcommand{\BibitemShut}[1]{}

\makeatother

\usepackage{babel}
\begin{document}
\global\long\def\E{\mathrm{e}}%
\global\long\def\D{\mathrm{d}}%
\global\long\def\I{\mathrm{i}}%
\global\long\def\mat#1{\mathsf{#1}}%
\global\long\def\vec#1{\mathsf{#1}}%
\global\long\def\cf{\textit{cf.}}%
\global\long\def\ie{\textit{i.e.}}%
\global\long\def\eg{\textit{e.g.}}%
\global\long\def\vs{\textit{vs.}}%
 
\global\long\def\ket#1{\left|#1\right\rangle }%

\global\long\def\etal{\textit{et al.}}%
\global\long\def\tr{\text{Tr}\,}%
 
\global\long\def\im{\text{Im}\,}%
 
\global\long\def\re{\text{Re}\,}%
 
\global\long\def\bra#1{\left\langle #1\right|}%
 
\global\long\def\braket#1#2{\left.\left\langle #1\right|#2\right\rangle }%
\global\long\def\obracket#1#2#3{\left\langle #1\right|#2\left|#3\right\rangle }%
 
\global\long\def\proj#1#2{\left.\left.\left|#1\right\rangle \right\langle #2\right|}%

\title{Spin transport in disordered long-range interacting spin chain}
\author{Benedikt Kloss}
\affiliation{Department of Chemistry, Columbia University, 3000 Broadway, New York,
New York 10027, USA}
\email{bk2576@columbia.edu}

\author{Yevgeny Bar Lev}
\affiliation{Department of Physics, Ben-Gurion University of the Negev, Beer-Sheva
84105, Israel}
\email{ybarlev@bgu.ac.il}

\begin{abstract}
Using a numerically exact technique we study spin transport and the
growth of an entanglement profiles in a disordered spin-chain with
long-range interactions, decaying as a power-law, $r^{-\alpha}$ with
distance and $1.75\leq\alpha\leq3.25$. Our study confirms the prediction
of recent theories that the system is delocalized in this parameters
regime. Moreover we find that for $\alpha>3/2$ the entanglement growth
is sublinear and the underlying transport is subdiffusive with a transient
super-diffusive tail. We show that an appropriately generalized Griffiths
picture shows diffusive transport and therefore does \emph{not} capture
the essential properties of the exact dynamics.
\end{abstract}
\maketitle
\emph{Introduction}.---Many-body localization (MBL) extends the notion
of Anderson localization to interacting systems \citep{Anderson1958b}.
For local interactions, its existence is well established theoretically
\citep{Basko2006a,Gornyi2005} and experimentally in one-dimensional
systems \citep{Schreiber2015a,Bordia2015,Smith2015} (see \citep{Abanin2017}
for a recent review), but there is evidence of localization also in
two-dimensional systems \citep{BarLev2015,Inglis2016a,Choi2016,Bordia2017,Kennes2018,Pupillo2019}.
For long-range interactions the fate of MBL is less clear. Some studies
suggest that the many-body localization is stable for $\alpha>2d$
\citep{Burin2006,Yao2014b,Gornyi2016,Tikhonov2018,Nag2019}, some
suggest it is stable for $\alpha>d/2$ \citep{Roy2019a}. Finite size
systems of size $L$ are claimed to exhibit an effective many-body-like
localization transition at a critical disorder, which for $\alpha<2d$,
scales like a power-law with the system size, and therefore diverges
in the thermodynamic limit \citep{Burin2006,Burin2015,Gornyi2016,Tikhonov2018,Gopalakrishnan2019}.
Some theories argue that delocalization occurs also for $\alpha>2d$
though with a critical disorder strength scales which increases slower
than algebraically with system size \citep{DeRoeck2016b,Tikhonov2018,Gopalakrishnan2019}.
Understanding the dynamics of disordered systems with long-range interactions
is of great importance to a number of physical systems, such as nuclear
spins \citep{Alvarez2015a}, dipole-dipole interactions of vibrational
modes \citep{Levitov1989,Levitov1990,Aleiner2011}, Frenkel excitons
\citep{Agranovich2008}, nitrogen vacancy centers in diamond \citep{Childress2006,Balasubramanian2009,Neumann2010,Weber2010,Dolde2013}
and polarons \citep{Alexandrov1996}. Long range interactions are
also common in atomic and molecular systems, where interactions can
be dipolar \citep{Saffman2010,Aikawa2012,Lu2012,Yan2013,Gunter2013,DePaz2013},
van der Waals like \citep{Saffman2010,Schauss2012}, or even of variable
range \citep{Britton2012,Islam2013,Richerme2014a,Jurcevic2014a}.
Some aspects of the dynamics in such systems were studied numerically
in Ref.~\citep{Safavi-Naini2018,Nag2019a,thomson2020localisation},
analytically in Ref.~ \citep{Gutman2015} and experimentally in Ref.~\citealp{Choi2018},
however spin transport in such systems was not considered.

The delocalized phase of one-dimensional systems with local interactions,
shows subdiffusive transport \citep{BarLev2014,Lev2014,Agarwal2014,Luitz2015a,Znidaric2016},
accompanied by sublinear growth of the entanglement entropy \citep{Luitz2017,Lezama2018,lezama_power-law_2019}
and intermediate statistics of eigenvalue spacing \citep{Serbyn2015}.
Anomalous transport is commonly explained by rare insulating regions,
which effectively suppress transport in one-dimensional systems. This
mechanism is known as the Griffith's picture \citep{Agarwal2014,Gopalakrishnan2015,Gopalakrishnan2015a}
(see Ref.~\citep{Agarwal2016_review} for a recent review and also
Ref.~\citep{DeRoeck2019a} were rare regions were introduced externally).
In dimensions higher than one the Griffiths picture predicts diffusion,
since rare regions can be circumvented \citep{Gopalakrishnan2015a},
however \emph{approximate} numerical studies \citep{BarLev2015} as
also recent experiments \citep{Choi2016,Bordia2017} suggest that
at least for short to intermediate times the relaxation and transport
appear to be anomalous. It is crucial to understand if this discrepancy
follows from incompleteness of the Griffiths picture or the approximation
of the method. While there are no efficient \emph{numerically exact}
methods to study the \emph{dynamics} of two-dimensional interacting
systems, some progress can be obtained for one-dimensional long-range
interacting systems. The Griffiths picture was not generalized to
this setting, but in analogy to the reasoning of higher dimensions
\citep{Gopalakrishnan2015a}, normal diffusive behavior is expected.

In a previous work we have shown that for \emph{clean} systems with
long-range interactions the local part of the Hamiltonian dictates
the spreading of the \emph{bulk} of a local spin excitation, while
the long-range part of the Hamiltonian only introduces a weak \emph{perturbative}
effect, in the face of power-law tails of the excitation profile,
with an exponent proportional to $\alpha$ \citep{Kloss2018a}. The
tails yield a superdiffusive signature of transport for all $\alpha$,
if a sufficiently high moment of the excitation profile is considered
\citep{Kloss2018a}. A natural question which arises is whether the
effect of long-range interactions in disordered systems goes beyond
a perturbative correction as it happens for their clean counterparts.
Moreover, if localization is destabilized by long-range interactions,
\emph{what is the resulting nature of spin transport}?

In this work we consider and answer these questions using a numerically
exact matrix product state (MPS) method. The study of long-ranged
interacting systems naturally requires large system sizes. In fact,
we show that for $\alpha=1.75$, finite size effects are pronounced
even for a chain of 51 spins, which is currently considered as the
state-of-the-art limit of exact diagonalization based techniques \citep{Wietek2018}.
MPS techniques are therefore indispensable to obtain numerically exact
results for chains with long-range interaction, albeit only up to
some finite time. This limitation arises since the required numerical
effort scales exponentially with the entanglement entropy of the state,
which for generic systems is known to grow linearly with time \citep{Kim2013}.
We stress that our aim here is \emph{not} to address the question
of stability of the MBL phase in the presence of long-range interactions,
but to study the dynamics in the delocalized phase. Moreover, since
is it technically hard to distinguish between very slow transport
and absence of transport, especially in a limited time-interval, our
method is not well suited for such purpose.

Time-evolution of long-ranged systems can be conveniently obtained
by the time-dependent variational principle (TDVP) applied to the
manifold of MPS \citep{Haegeman2011,Haegeman2013,Haegeman2016}. It
was successfully utilized to study the dynamics of spin chains with
local interactions in disordered or quasiperiodic potentials \citep{Doggen2018,Doggen2019}.
For low bond dimensions and very far from the numerically exact limit
this method was proposed as an inexpensive candidate to achieve accurate
hydrodynamic description of transport \citep{Leviatan2017}, however
it was shown to be unreliable for generic systems \citep{Kloss2017}.
In this study, we use TDVP as a \emph{numerically exact} method, and
study the nature of transport in long-range-interacting disordered
one-dimensional spin chain. We focus on parameter regimes in which
the interaction is sufficiently short-ranged such that the corresponding
clean system shows asymptotic diffusive behavior, and disorder ranges
for which the system is argued to be delocalized by all existing theories.

\emph{Model}.--- We study transport properties of the one-dimensional
long-ranged disordered Heisenberg model,
\begin{equation}
\hat{H}=J\sum_{i=1}^{L-1}\sum_{j>i}^{L}\frac{1}{\left(j-i\right){}^{\alpha}}\left(\hat{S}_{i}^{x}\hat{S}_{j}^{x}+\hat{S}_{i}^{y}\hat{S}_{j}^{y}+\hat{S}_{i}^{z}\hat{S}_{j}^{z}\right)+\sum_{i=1}^{L}h_{i}\hat{S}_{i}^{z},\label{eq:XXZ-Ham}
\end{equation}
where $h_{i}$ is uniformly distributed in the interval $\left[-W,W\right]$
and $\hat{S}_{i}^{\left(x,y,z\right)}$ are the appropriate projections
of the spin-$1/2$ operators on site $i$. In the following, we use
$J=1$, which sets the time unit of the problem. To study the dynamical
properties of this model we start the system from a random product
state, $\ket n$, in the eigenbasis of $\hat{S}_{i}^{z}$ and calculate
the growth of the bipartite entanglement entropy $S\left(t\right)$
as also the spreading of a spin-excitation as a function of time,
which is assessed from the two-point spin correlation function,
\begin{equation}
C_{x}^{n}(t)=\left\langle n\left|\hat{S}_{L/2+x}^{z}\left(t\right)\hat{S}_{L/2}^{z}\left(0\right)\right|n\right\rangle .\label{eq:profile}
\end{equation}
Entanglement entropy is directly available since we use the two-site
TDVP method \citep{Haegeman2016}. We then average both $S\left(t\right)$
and $C_{x}^{n}\left(t\right)$, by randomly sampling both the disorder
and the initial state of the system, such that any state has an equal
probability to occur. This corresponds to infinite temperature ensemble,
$C_{x}\left(t\right)=\mathcal{N}^{-1}\sum C_{x}^{n}\left(t\right)$,
where $\mathcal{N}$ is the Hilbert space dimension. It is convenient
to characterize the spreading by the width of the averaged excitation
profile, 
\begin{equation}
\sigma^{2}\left(t\right)=\sum_{x=-L/2}^{L/2}x{}^{2}C_{x}\left(t\right)\label{eq:MSD}
\end{equation}
 which is analogous to the classical mean-square displacement (MSD).
Typically the MSD scales as, $\sigma^{2}(t)\sim t^{\gamma}$, with
$\gamma=2$ ($\gamma=1$) for ballistic (diffusive) transport and
$0<\gamma<1$ corresponding to subdiffusive transport. We use the
log-derivate to define a time-dependent dynamical exponent $\gamma\left(t\right)=\mathrm{d}\ln\sigma^{2}\left(t\right)/\mathrm{d}\ln t$,
which asymptotically converges to $\gamma.$ We similarly define the
time-dependent dynamical exponent $\delta\left(t\right)=\mathrm{d}\ln S\left(t\right)/\mathrm{d}\ln t$,
to characterize the spread of the entanglement.

\emph{Method}.--- The Hilbert-space dimension of a quantum lattice
systems scales exponentially with the size of the system. Any wavefunction
in the Hilbert space can be written as a matrix product state (MPS),

\begin{equation}
\ket{\Psi[A]}=\sum_{\{s_{n}\}=1}^{d}A^{s_{1}}(1)A^{s_{2}}(2)\dots A^{s_{N}}(L)\ket{s_{1}s_{2}\dots s_{L}}\label{eq:MPS-gen}
\end{equation}
where $d$ is the local Hilbert space dimension, $A^{s_{i}}(i)\in\mathbb{C}^{D_{i-1}\times D_{i}}$
are complex matrices and $D_{0}=D_{L}=1$, such that the product of
matrices evaluates to a scalar coefficient for a given configuration
$\ket{s_{1}s_{2}\dots s_{L}}$. To be an \emph{exact} representation
of the wavefunction the dimension of the matrices, the bond dimension,
must scale exponentially with the systems size. Typically one approximates
the wavefunction by truncating the dimension of the matrices to a
predetermined dimension with computationally tractable number of parameters.
Exact results are obtained when the approximate dynamics are converged
with respect to the bond dimension.

The time-dependent variational principle (TDVP) allows one to obtain
a locally optimal (in time) evolution of the wavefunction on the manifold
of MPS, $\mathcal{M}_{\chi},$ with some fixed bond dimension $\chi$.
It amounts to solving a tangent-space projected Schrödinger equation
\citep{Haegeman2016}:
\begin{equation}
\frac{d\ket{\Psi[A]}}{dt}=-\I P_{\mathcal{M}}\hat{H}\ket{\Psi[A]},\label{eq:TDVP}
\end{equation}
where $P_{\mathcal{M}}$ is the tangent space projector to the manifold
$\mathcal{M}_{\chi}$.\textcolor{magenta}{{} }Equation \eqref{eq:TDVP}
is solved using a second-order Trotter-Suzuki decomposition of the
projector. The Hamiltonian is approximated as a sum of exponentials,
which can be efficiently represented as an MPO \citep{Crosswhite2008a}.
The number of exponentials is chosen such that the resulting couplings
do not differ by more than 2\% from the exact couplings for any pair
of sites. Through this work we have used a bond-dimension of up to
$\chi=1024$ and timestep of $dt=0.1$ and verified that our results
are convergent with respect to these numerical parameters (see \citep{SuppMat2019}).
We average over initial conditions and disorder realizations at the
same time and use 1000 realizations unless stated otherwise. All calculations
are performed using the TenPy library using a two-site version of
the TDVP for MPS and exploiting that the Hamiltonian conserves the
total magnetization \citep{Hauschild2018}, which allows us to directly
access to the growth of the entanglement entropy.

\emph{Results}.---

\begin{figure}
\includegraphics{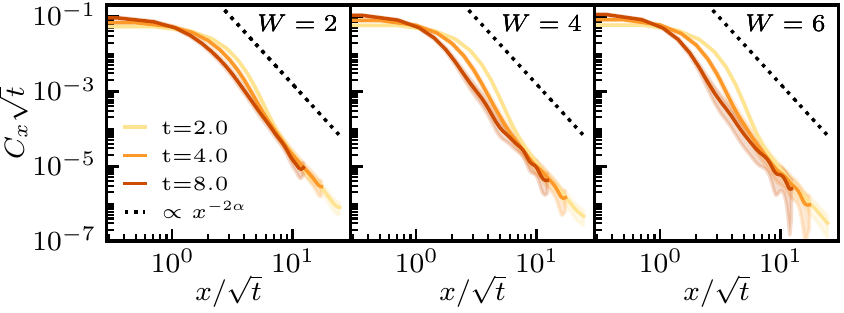}

\caption{\label{fig:profiles}Rescaled magnetization profiles at different
times on log-log scale for bond dimension $\chi=512$ and different
disorder strengths ($W=2.0,4.0$ and $6.0$ from left to right). The
shaded area shows the standard deviation of the profile obtained from
a bootstrapping procedure. Profiles are smoothed by a Gaussian filter
with a standard deviation of $2.0$. Black dotted line is a guide
to the eye of a power-law, $x^{-2\alpha}$.}
\end{figure}
\begin{figure}
\centering{}\includegraphics[width=1\columnwidth]{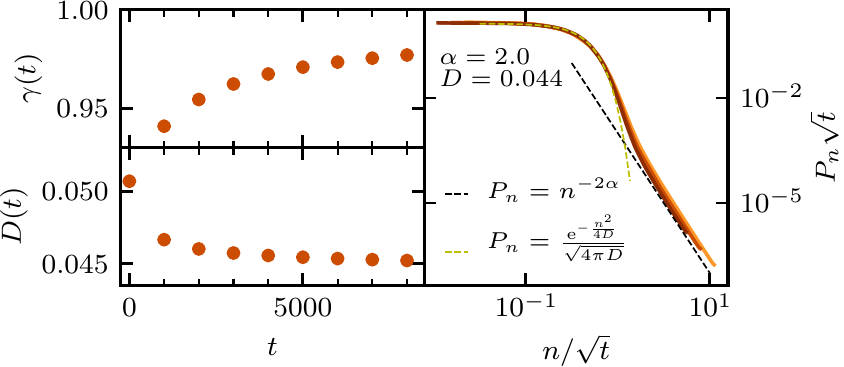}\caption{\label{fig:griffiths-msd-gamma}\emph{Right panel}. A rescaled log-log
plot of $P_{n}(t)$ for $\alpha=2.0$, $h_{0}=8$ at various times
$t$, computed from the solution of the generalized Griffiths model
(\ref{eq:longrange_master}). \emph{Left panel}s. The corresponding
dynamical exponent $\gamma\left(t\right)$ (top panel) and $D(t)$
(bottom panel).}
\end{figure}

For transport that is not purely diffusive, the MSD contains only
\emph{partial} information on transport, since in this case the asymptotic
shape of the profile is \emph{not} described by a Gaussian. To get
a full picture of transport it is therefore pertinent to examine the
evolution in time of the excitation profiles, which we do in Fig.~\ref{fig:profiles}.
Similarly to the clean case in Refs.~\citealp{Kloss2018a,Schuckert2019a},
the tails of the excitation profile follow a power-law of $-2\alpha$
regardless of the disorder strength, which shows that the disorder
cannot suppress the long-range hops of the spin. These tails are responsible
for the divergence of the MSD with system size for $\alpha<3/2$.
The failure of the rescaling procedure performed in Fig.~\ref{fig:profiles},
which is expected to yield a perfect collapse for diffusive transport
(c.f. Fig.~\ref{fig:griffiths-msd-gamma}), indicates that transport
is \emph{not} diffusive.

To assess the influence of the disorder on the dynamics we focus on
$\alpha=1.75$ and compute the averaged bipartite entanglement and
the MSD (\ref{eq:MSD}) for a number of disorder strength $W\in\left[2,12\right],$which
are predicted to be in the delocalized phase\citep{Tikhonov2018}{]}.
\begin{figure}
\includegraphics{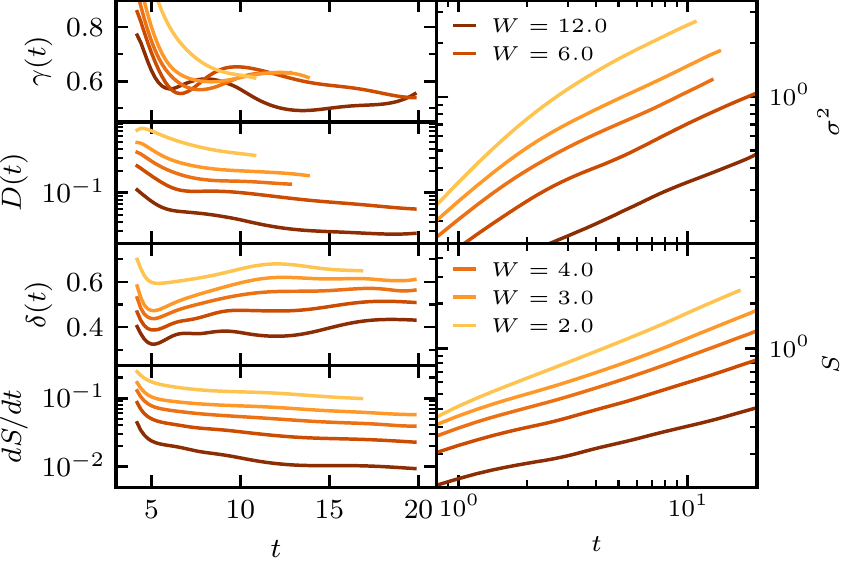}

\caption{\label{fig:MSD} \emph{Right column}. MSD (top panel) and entanglement
entropy $S(t)$ (bottom panel) on log-log scale after numerical filtering
(see main text) as a function of time for different disorder strengths
($W=2-12$) and $\chi=512$. \emph{Left column}. $\gamma\left(t\right)$
and $D\left(t\right)$ computed from filtered MSD data, smoothed with
a moving average of width $t=4$ (two upper panels), and similarly
$\delta\left(t\right)$ and $dS/dt$ computed from filtered $S\left(t\right)$.}
\end{figure}
Figure \ref{fig:MSD} shows the MSD (\ref{eq:MSD}) and the entanglement
entropy, $S(t)$, as a function of time for various disorder strengths
together with the corresponding linear and logarithmic derivatives.
All data is converged with respect to the system size ($L=75)$ except
for the weakest disorder strength $\left(W=2\right)$ (see \citep{SuppMat2019}).
At strong disorder, oscillatory features emerge, with a period of
the order of the hopping rate. These oscillations are common in disordered
systems, and typically correspond to oscillations between nearby localization
centers. Such oscillations hinder to reliably extract the dynamical
exponent. In order to rectify this issue we filter-out the corresponding
frequency in the Fourier domain (for raw data and a description of
the procedure see \citep{SuppMat2019}). As can be seen from Fig.~\ref{fig:MSD}
the linear derivatives of MSD and $S\left(t\right)$ are monotonically
decreasing with time, while their log-derivatives appear to converge
to a constant value smaller than 1. This observation points towards
a \emph{sub-linear} dependence of MSD and $S\left(t\right)$, which
is indicative of \emph{subdiffusive} transport.

\begin{figure}
\includegraphics{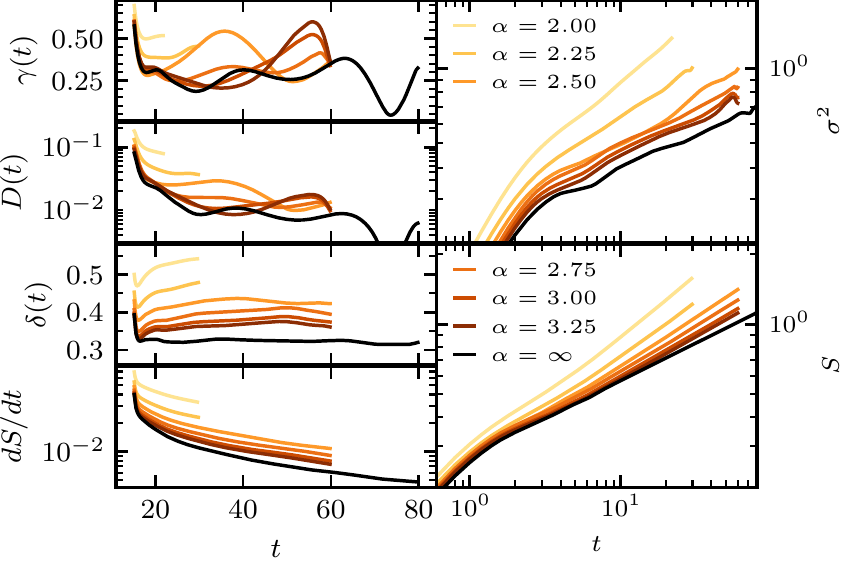}

\caption{\label{fig:MSD-S-alpha}Similar to Fig.~\ref{eq:MSD}, but for $W=3.0$
and $2.0\protect\leq\alpha\protect\leq3.25$. System sizes are $L=75$
for $\alpha\protect\leq2.5$, $L=51$ for $\alpha>2.5$ and $L=35$
for local interactions.}
\end{figure}

In Fig.~\ref{fig:MSD-S-alpha} we examine the dependence of the dynamics
on the range of the interaction by fixing the disorder strength $\left(W=3.0\right)$
and varying $2.0\leq\alpha\leq3.25$. The disorder is chosen, such
that in the local limit, $\alpha\to\infty$ (black line in Fig.~\ref{fig:MSD-S-alpha}),
the system is delocalized and subdiffusive \citep{Lev2014}. Similarly
to Fig.~\ref{eq:MSD}, the dynamical exponents $\gamma$ and $\delta$
converge to a constant value smaller than 1 for all studied $\alpha$'s,
which is monotonically decreasing with $\alpha$. 
\begin{figure}
\includegraphics{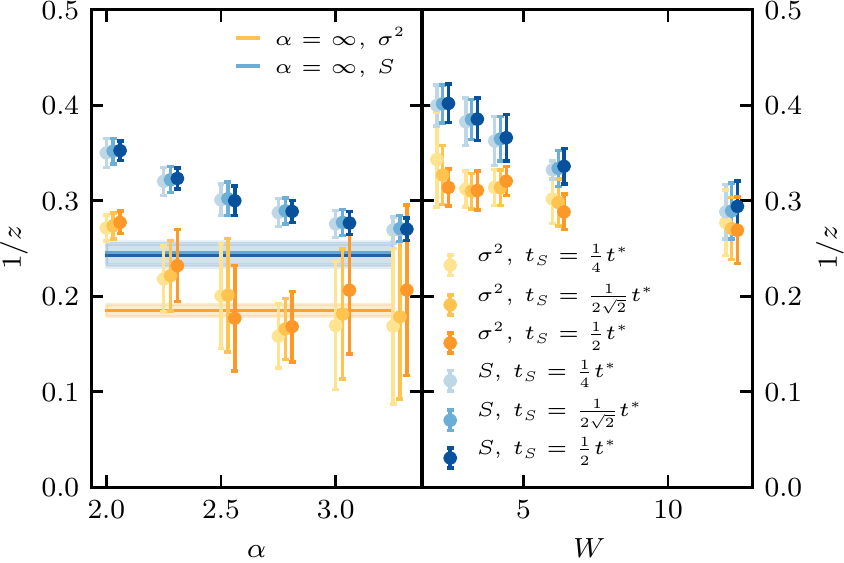}

\caption{\label{fig:dynamical-exp-alpha}Dynamical exponent $1/z$ for MSD
(orange hues) and $S(t)$ (blue hues) as a function of $\alpha$ for
$W=3.0$ (left panel) and as a function of $W$ for $\alpha=1.75$
(right panel). The exponents and the error bars are obtained as average
and standard deviation of the filtered data for the MSD and $S\left(t\right)$
over different windows $\left[t_{S},2^{-\frac{1}{4}}t^{*}\right]$
for left panel and $\left[t_{S},t^{*}\right]$ for right panel, where
$t^{*}$ is the time up to which data is converged.}
\end{figure}
In Fig.~\ref{fig:dynamical-exp-alpha} we plot the dynamical exponents
as extracted from Figs.~\ref{fig:MSD}, \ref{fig:MSD-S-alpha} for
different $W$s and $\alpha$'s. We use the relation between the exponents
proposed in Refs.~\citep{Vosk2014,Potter2015} (see also Ref.~\citep{Luitz2016c}),
$1/z=\gamma/2=\delta/\left(1+\delta\right)$ . While the relation
is not satisfied well, the overall dependence of the exponents appear
similar, with both exponents monotonically decreasing with $\alpha$
and converging to the $\alpha\to\infty$ limit. For $S\left(t\right)$
the dynamical exponent is reliable in the entire range of parameters,
but the oscillations in the MSD result in large error bars for $\alpha>2.5$.

Since the exact numerical study of long-range systems is rather limited
in time, it is beneficial to find a phenomenological model which attempts
to reproduce the relevant dynamical features, and at the same time
suggests an effective mechanism. For disordered \emph{local} systems
the Griffiths picture serves this purpose \citep{Agarwal2014,Gopalakrishnan2015,Gopalakrishnan2015a}.
We generalize the Griffiths picture to long-range systems by introducing
a finite probability for long jumps with a rate which decays as, $x^{-2\alpha}$,
in accord with the long-range part of the Hamiltonian \footnote{The factor of 2 stems from the difference between amplitude and probability}.
This reduces to the following master equation,
\begin{align}
\frac{\partial P_{n}}{\partial t} & =\sum_{i}W_{in}P_{i}-\left(\sum_{j}W_{nj}\right)P_{n}\label{eq:longrange_master}\\
W_{ij} & =\frac{\E^{-h_{ij}}}{\left|i-j\right|^{2\alpha}}\qquad i\neq j\nonumber 
\end{align}
where $h_{ij}$ is a symmetric matrix composed of independent random
variables, which stand for the heights of the barriers. The precise
shape of the distribution of the barrier heights is not important,
as long as it is unbounded, guaranteeing the existence of very weak
links. To be concrete, we take it to be the exponential distribution,
$p\left(h\right)=h_{0}^{-1}\exp\left[-h/h_{0}\right]$. We note in
passing, that while the form of the transition matrix is similar to
the power-law random banded matrices used to study Anderson localization
with power-law hopping \citep{Mirlin1996}, there are crucial differences:
(a) we are applying it to a \emph{classical} problem, (b) $W_{ij}$
has many-elements close to zero, and must satisfy, $W_{ii}=-\sum_{i\neq j}W_{ij}$.
Since long-hops effectively avoid weak-links, such model is expected
to be diffusive, but it is important to see how it approaches diffusion
as a function of time. To examine that, we numerically solve (\ref{eq:longrange_master})
for about 500 realizations of the transitions rate matrix, $W_{ij}$,
with $h_{0}=8$ and a lattice size of $L=1000$. At time $t=0$ the
walker is initiated at the origin, $P_{n}\left(t=0\right)=\delta_{n0}$.
The probability to find a walker at site $n$ for various times has
a Gaussian form in the bulk, followed by a power-law tail, which can
be better seen after the rescaling, $\sqrt{t}P_{n}\left(n/\sqrt{t}\right)$
(see Fig.~\ref{fig:griffiths-msd-gamma}). Thus transport in this
model is asymptotically diffusive. In Fig\@.~\ref{fig:griffiths-msd-gamma}
we show the $D\left(t\right)$ and $\gamma\left(t\right)$ for this
generalized Griffiths model as a function of time. We note that while
$\gamma\left(t\right)$ converges to 1 and $D\left(t\right)$ converges
to a constant, indicative of diffusive transport the convergence is
quite slow.

\emph{Discussion}.---Using a numerically exact method we studied
transport in a disordered spin-chain with interactions between the
spins decaying as $x^{-\alpha}$ with distance. For \emph{clean} systems
and $\alpha>d/2$ the long-range interaction appears to be a \emph{perturbative}
effect which is manifested by power-law tails of the relevant excitation
profile, while the dynamics of the bulk of the excitation is governed
by the local interactions \citep{Kloss2018a,Luitz2018}. Carrying
over this analysis to long-range disordered systems suggests localization
for sufficiently strong disorder, since \emph{local} interacting systems
exhibit MBL. On the other hand, most studies argue that the addition
of long-range interactions is not perturbative, and leads to destruction
of localization due to a prevalence of resonances \citep{Burin2006,Yao2014b,Gornyi2016,Tikhonov2018}.
Our study shows that in accord with these works even in the presence
of the strongest disorder the system is delocalized. The long-range
part of the Hamiltonian thus destabilizes localization and leads to
spreading of spin excitations. Our results are consistent with subdiffusive
transport, which can be seen in a sub-linear growth of the bipartite
entanglement and the MSD. As with any numerical result, our analysis
is based on \emph{finite} times data, therefore we cannot rule out
slow drift towards diffusion which might occur for times inaccessible
to numerically exact studies. Our findings are \emph{not} consistent
with the prediction for finite heat conductivity, $\kappa\sim W^{-3},$which
can be obtained from Ref.~\citep{Gutman2015} for the heat conductivity
by setting $\alpha=1.75$ and the temperature to $T=W$. Furthermore,
we find that the scaling of the critical disorder strength with system
size, advocated by existing theories, does not affect transport in
the delocalized phase, at least far away from the critical point.
The observed subdiffusive transport is also \emph{not} consistent
with a generalization of the Griffiths mechanism \citep{Gopalakrishnan2015a},
which predicts diffusion for long-range systems, since long-range
interactions introduce an effective way to circumvent blocking regions.
We have confirmed this by a numerical solution of a long-range random-barrier
model (\ref{eq:longrange_master}). While in this model, the convergence
of the diffusion coefficient to its asymptotic value is quite slow,
unlike the long-range model (\ref{eq:XXZ-Ham}), the dynamical exponent
is very close to 1, already for short-times, and doesn't exhibit any
plateau for all times. This finding adds up to the mounting evidence
against the importance of bottle-necks for subdiffusion, as was shown
already in Refs.~\citep{BarLev2015,BarLev2017a} (cf. \citep{Znidaric2018a,Varma2019b},
and see also the very recent \citep{Schulz2019}), suggesting that
our understanding of the mechanism of anomalous transport in the vicinity
of the MBL transition is far from being complete.
\begin{acknowledgments}
The authors acknowledge fruitful discussions with Alexander Burin.
This research was supported by the Israel Science Foundation (grants
No. 527/19 and 218/19). BK acknowledges funding through the National
Science Foundation Grant No. CHE-1464802.
\end{acknowledgments}

\subsection{Supplementary material}

\begin{figure}
\includegraphics{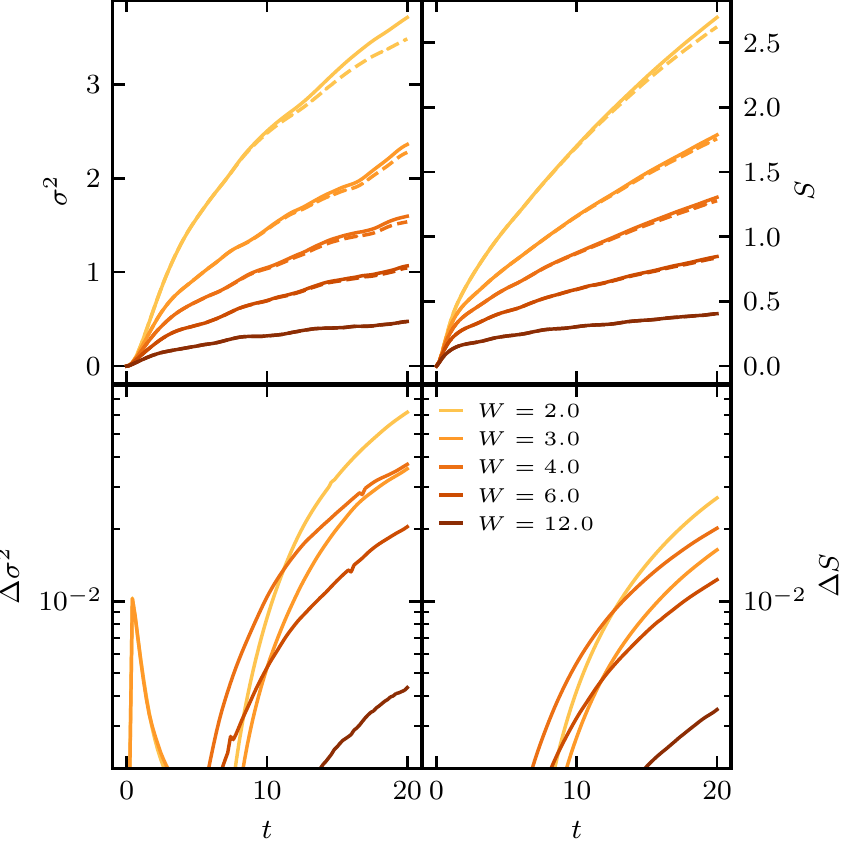}

\caption{\label{fig:conv-MSD-chi}Convergence of the MSD and $S\left(t\right)$
with respect to the bond dimension at $\alpha=1.75$ for various disorder
strengths $W$. \emph{Upper panel}: MSD and $S\left(t\right)$ at
reference bond dimensions $\chi=1024$ ($W=2$ and 3, solid) and $\chi=512$
($W=4,6$ and 12 solid) and half the reference bond dimension (dashed).
\emph{Lower panel}: Relative errors $\Delta\sigma^{2}$(left panel)
and $\Delta S$ (right) between calculations at the reference and
half the reference bond dimension. The system size for all panels
is $L=75$ and a time-step of $dt=0.1$ was used.}
\end{figure}
\begin{figure}
\includegraphics{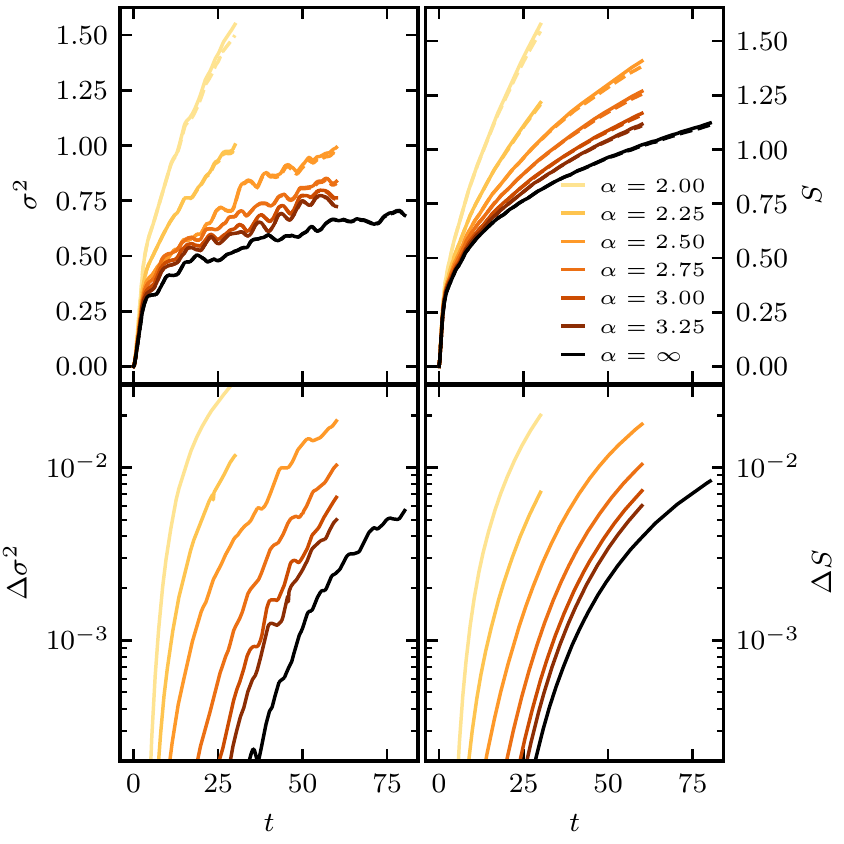}

\caption{\label{fig:conv-MSD-chi-1}Convergence of the MSD and $S\left(t\right)$
with respect to bond dimension for various $\alpha$ and a disorder
strength of $W=3.0$. \emph{Upper panel}: MSD and $S\left(t\right)$
at bond dimensions $\chi=512$ (solid) and $\chi=256$ (dashed). \emph{Lower
panel}: Relative deviation $\Delta\sigma^{2}$ (left panel) and $\Delta S$
(right) of calculations with $\chi=256$ and $\chi=512$.. The system
sizes for all panels are $L=75$ for $\alpha\protect\leq2.5$, $L=51$
for $2.5<\alpha\protect\leq3.25$ and $L=35$ for $\alpha=\infty$.
A time-step of $dt=0.1$ was used.}
\end{figure}
\begin{figure}
\includegraphics{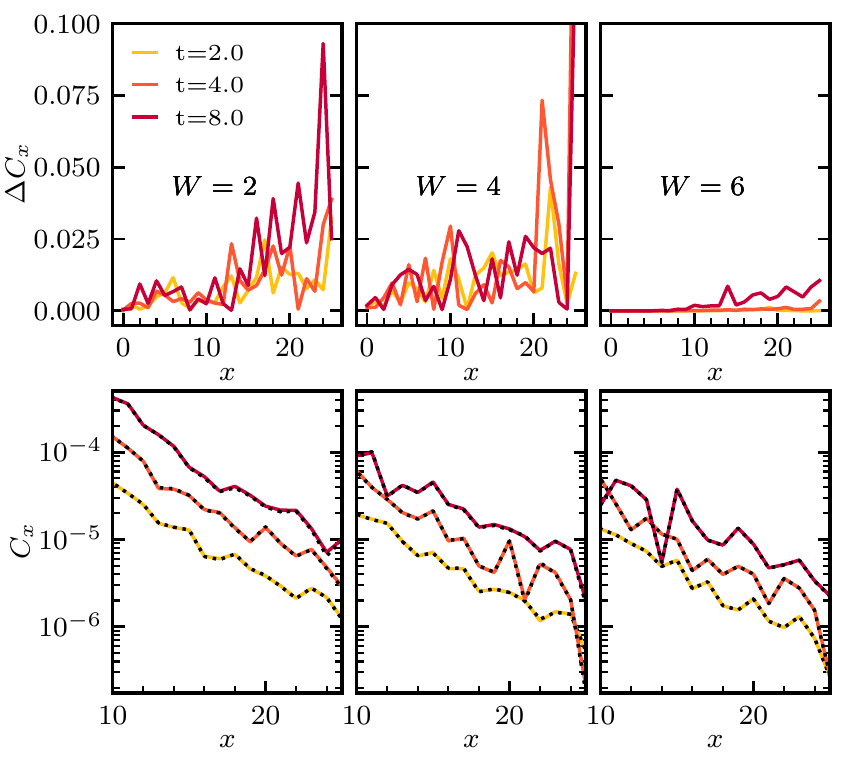}\caption{\label{fig:conv-tails}Convergence of the spin excitation profiles
with respect to bond dimension for $L=51$ and $dt=0.1$ at various
disorder strengths and times. \emph{Upper panels}: Relative error
$\Delta C_{x}$ between calculations with $\chi=512$ and $\chi=256$.
\emph{Lower panels}: Tails of spin excitation profiles with $\chi=512$
(solid lines) and $\chi=256$ (black dotted lines).}
\end{figure}

\emph{Convergence with respect to numerical parameters}.---Numerical
exactness of the dynamics generated by TDVP-MPS is obtained by converging
with respect to the bond-dimension, $\chi$, as well as the time-step,
$dt$. In Fig.~\ref{fig:conv-MSD-chi} , we provide comparisons of
the mean-square displacement (MSD) and the entanglement entropy $S(t)$
from calculations with bond-dimensions up to $\chi=1024$. All results
for the MSD and $S(t)$ reported in the main article are converged
up to a deviation of 2~\% between the two largest bond dimensions.
In order to check for convergence with the time-step it is sufficient
to use a smaller bond dimension, since time-step errors are usually
more severe at smaller bond dimension. 
\begin{figure}
\includegraphics{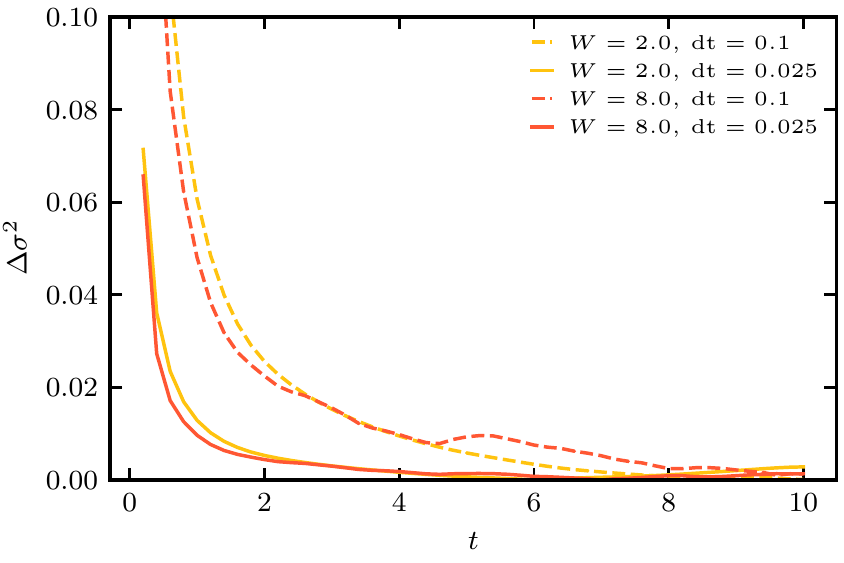}

\caption{\label{fig:conv-MSD-dt}Convergence of the MSD with respect to time-step.
Relative error $\Delta\sigma^{2}$ compared to a reference calculation
with time-step $dt=0.005$ for $L=51$, $\chi=64$ at weak and strong
disorder.}
\end{figure}
Fig.~\ref{fig:conv-MSD-dt} shows the relative deviation of the MSD
at several time-steps from a reference calculation at time-step $dt=0.005$.
The large relative error initially is caused by an approximately constant
error in absolute terms and that becomes negligible in relative terms
after times larger than a few units of the hopping. A time-step of
$dt=0.1$ is thus sufficient to obtain a converged MSD within the
range of disorder strengths studied. Evaluating the spatial spin excitation
profile in the tails at strong disorder becomes sensitive to numerical
noise for small values of the correlation function, $C_{x}$, and
is limited by a complex interplay of time-step errors and accumulation
of numerical round-off errors. As shown in Fig.~\ref{fig:conv-tails},
the convergence of the tails of the spin excitation profile with respect
to bond dimension is generally well controlled ($<5\%)$ up to times
for which the MSD is converged as well.

\begin{figure}
\includegraphics{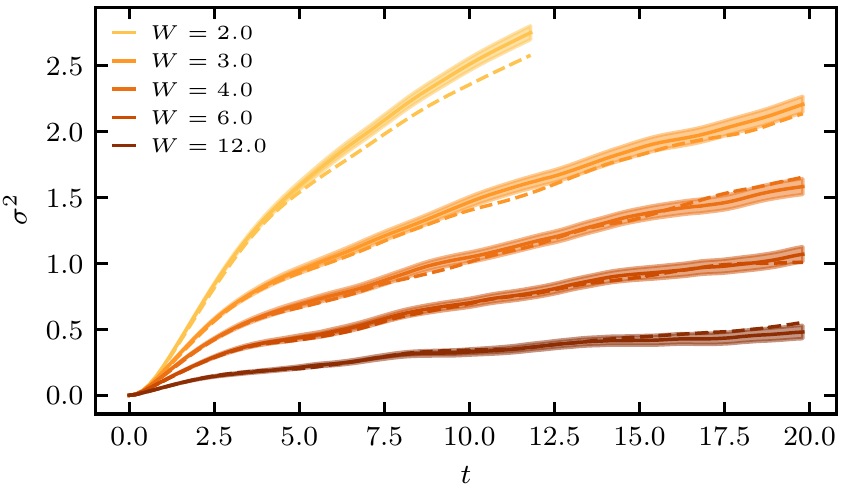}\caption{\label{fig:finite-size}Comparison of the MSD for various disorder
strengths at $\alpha=1.75$ for system sizes $L=75$ (solid lines)
and $L=51$ (dashed lines). The shaded area indicates the standard
deviation for $L=75$ obtained from bootstrap sampling.}
\end{figure}

\emph{Finite size effects}.---In Fig.\,\ref{fig:finite-size} we
provide evidence that the MSD is converged with respect to system
size for $L=75$ for the data presented in the main article at $\alpha=1.75$
for all but the smallest disorder strengths $W=2.0$.

\begin{figure}
\includegraphics{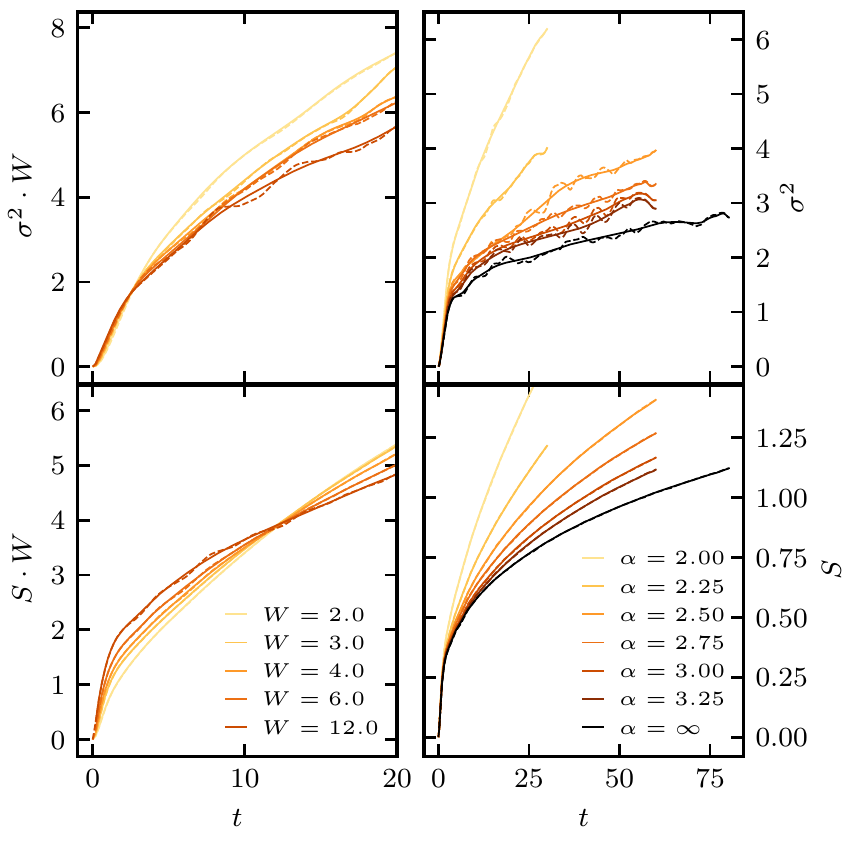}\caption{\label{fig:filtering}Comparison of filtered (solid lines) and unfiltered
(dashed lines) MSD (upper panels) and entanglement entropy $S\left(t\right)$
(lower panels) for various disorder strengths at $\alpha=1.75$ (left
panels) and for various $\alpha$ with a disorder strength of $W=3.0$
(right panels). For improved visibility, the data for $\alpha=1.75$
(left panels) is rescaled with the disorder strength.}
\end{figure}

\emph{Filtering out high-frequency oscillations}.---The presence
of strong disorder leads to high frequency oscillations which average
only slowly and are an obstacle in analyzing transport properties
quantitatively. Hence, we smooth our data by removing the high frequency
oscillations according to the following protocol. The linear time
derivative of the data is Fourier transformed and a Gaussian broadening
is applied in the Fourier domain before transforming back to the time
domain, from which the filtered data is obtained by integration. We
find that applying a weak broadening at low frequencies, and successively
increasing the strength of the broadening at higher frequencies results
in an efficient and unbiased removal of the high frequency oscillations.
First a broadening of width $w=0.25$ is applied to the range of all
nonzero frequencies, followed by a broadening of width $w=0.75$ excluding
the two lowest frequencies, and finally a broadening of width $w=1.5$
applied to all but the 4 lowest frequencies. We note that the result
depends weakly on the exact values of these parameters. This processing
does not result in a systematic bias compared to the raw data, as
shown in Fig.\,\ref{fig:filtering}. The smoothing becomes inefficient
towards the boundaries of the support of the data in the time-domain.
When available, the raw data has been used past its convergence time
as an input for the filtering to circumvent this problem. In the main
text, we report only the filtered data and only up to the convergence
time determined from Figs.~\ref{fig:conv-MSD-chi} and \ref{fig:conv-MSD-chi-1}.
While this can in principle introduce a bias for the filtering at
late times, we verified that the filtered data is consistent with
the raw data for all converged times.

\bibliographystyle{apsrev4-1}
\bibliography{lib_yevgeny,local}

\end{document}